\documentclass[%
 reprint,
 amsmath,amssymb,
 aps,
]{revtex4-1}

\usepackage{graphicx}
\usepackage{dcolumn}
\usepackage{bm}

\usepackage{color}

\newcommand{\er}[1]{Eq.~\eqref{#1}}
\newcommand{\ers}[2]{Eqs.~(\ref{#1}-\ref{#2})}

\begin{document}


\title{
Applicability of dynamic facilitation theory to binary hard disk systems
}

\author{Masaharu Isobe}
\email{isobe@nitech.ac.jp}
\affiliation{Graduate School of Engineering, Nagoya Institute of Technology,
Nagoya, 466-8555, Japan}

\author{Aaron S. Keys}
\affiliation{Department of Chemistry, University of California, Berkeley, California 94720, USA}

\author{David Chandler}
\affiliation{Department of Chemistry, University of California, Berkeley, California 94720, USA}

\author{Juan P. Garrahan}
\affiliation{School of Physics and Astronomy, University of Nottingham, Nottingham NG7 2RD, UK}

\date{\today}

\begin{abstract}
We investigate numerically the applicability of dynamic facilitation (DF) theory for glass-forming binary hard disk systems where supercompression is controlled by pressure.  By using novel efficient algorithms for hard disks, we are able to generate equilibrium supercompressed states in an additive non-equimolar binary mixture, where micro-crystallization and size segregation do not emerge at high average packing fractions.  Above an onset pressure where collective heterogeneous relaxation sets in, we find that relaxation times are well described by a ``parabolic law'' with pressure.  We identify excitations, or soft-spots, that give rise to structural relaxation, and find that they are spatially localized, their average concentration decays exponentially with pressure, and their associated energy scale is logarithmic in the excitation size.  These observations are consistent with the predictions of DF generalized to systems controlled by pressure rather than temperature.
\end{abstract}


\maketitle


It is highly debated which of the competing theoretical approaches to the glass transition is the most appropriate to describe the relaxational dynamics of glass formers \cite{Ediger1996, Kob2005, Lubchenko2007, Chandler2010, Berthier2011, Biroli2013}.  One  perspective is provided by so-called dynamic facilitation (DF) theory \cite{Chandler2010} which is based on the detailed study \cite{Garrahan2002,Garrahan2003,Merolle2005,Garrahan2007} of idealized kinetically constrained models (KCMs) \cite{Ritort2003}.  The central predictions from DF are: (i) In the supercooled regime relaxation originates from localized excitations or soft-spots distributed randomly with a concentration that decreases exponentially with inverse temperature; their kinetics is facilitated (excitations allow for relaxation in their vicinity), giving rise to heterogeneous
  dynamics \cite{Chandler2010}.  Excitations are explicit in KCMs, but in actual glass formers they would be emergent \cite{Chandler2010}.  (ii) Relaxation is ``hierarchical'' (as in the East facilitated model \cite{Jackle1991,Sollich1999} or its generalisations \cite{Ritort2003}) leading to an overall relaxation time that follows a ``parabolic'' law \cite{Elmatad2009}, i.e. the exponential of a quadratic function of inverse temperature, which while super-Arrhenius is distinct from the empirical V\"ogel-Fulcher-Tammann (VFT) law \cite{Ediger1996,Kob2005,Lubchenko2007,Chandler2010,Berthier2011,Biroli2013}, in particular as it has no finite temperature singularity.  (iii) Underlying glassy slowing down is a non-equilibrium ``space-time'' transition \cite{Merolle2005, Garrahan2007} whose fluctuations manifest as dynamic heterogeneity.  

The above predictions of DF have been seen to hold in {\em thermal} atomistic systems: Effective excitations can be identified  \cite{Keys2011,Speck2012} through path sampling techniques \cite{dellago_2002,bolhuis_2002}, and are found to conform to (i) above; the parabolic law (ii) is an adequate description of relaxation rates at low temperature for experimental liquids \cite{Elmatad2009}; and active/inactive transitions (iii) are indeed found in simulations of atomistic liquids by means of large deviation techniques \cite{hedges_2009, Speck2012b}. 
Here, we extend the DF approach to systems where the controlling parameter is pressure, exploring in detail the validity of predictions (i) and (ii) for binary hard disks.  While there can be differences in specific aspects of the dynamics between dense systems in dimensions two and three~\cite{Flenner2015} (and even more significantly on their thermodynamics), we are interested here in general properties of slow  relaxation under supercompressed conditions - two-dimensional systems are a useful test ground as they can be studied exhaustively, as we describe below.  

In systems of hard particles, the primary control parameter is the either the pressure $p$ or the packing fraction $\nu$.  In analogy with the thermal problem, we will consider the case where pressure is the control parameter.  We make natural extensions of the basic DF scaling relations. First, prediction (i) implies that the density of excitations $c_{a}$, where $a$ is the size of particle displacement used to identify an excitation (see below), goes as 
\begin{equation}
c_a \propto \exp{[-\kappa_a (p^*-p^*_0)]} .  
\label{eqn:edf1}
\end{equation} 
Here $p^*$ is the reduced pressure $p^* = \beta p \sigma_*^2$, where $\sigma_*$ is the effective diameter in a binary mixture~\cite{eff_d}, with $\beta = 1/k_{\rm B} T$ and $k_{\rm B}$ the Boltzmann constant.  In \er{eqn:edf1} $p^*_0$ is the {\em onset pressure} above which cooperative and heterogeneous dynamics becomes significant (setting the regime of validity of the DF approach). 
Furthermore, the scale $\kappa_a$ associated to a displacement of size $a$ should grow logarithmically as,
\begin{equation}
\kappa_a - \kappa_{a^{'}} = \gamma \kappa_{\sigma_{*}} \ln{(a/a^{'})} \label{eqn:edf2} 
\end{equation}
\noindent
where $\gamma$ is a non-univeral exponent of order unity, and 
$\kappa_{a^{'}}$ and $\kappa_{\sigma_{*}}$ are $\kappa_a$ at $a=a^{'}$ and $a=\sigma_*$, respectively. 
Similarly, for (ii) we have that 
the primary relaxation time, $\tau_{\alpha}$, should be the exponential of a parabolic function of the reduced pressure,
\begin{equation}
\tau_\alpha = \tau_{0} \exp{[\kappa^2(p^*-p^*_0)^2]} ,  \label{eqn:edf3}
\end{equation} 
with $\kappa$ a system specific activation scale. 
\ers{eqn:edf1}{eqn:edf3} are the equivalents of the DF scaling forms to those applicable to soft interactions, see Refs.\ \cite{Elmatad2009,Keys2011}.

We consider $N$ additive binary hard disks, 
of small and large diameters $\sigma_0$ and $\sigma_1$, and mole fractions $x_0$ and $x_1$, respectively. The size ratio is $\alpha=\sigma_1/\sigma_0$, and the system is in a $L_x \times L_y$($=A$) square box ($L_y/L_x = 1$) with periodic boundaries.  We aim to establish, at high densities, {\em true equilibrium} where the equilibrium distribution can be reached from any initial condition.
Given that structural relaxation becomes extremely slow at high densities, to accelerate equilibration we use an efficient algorithm based on Event-Chain Monte Carlo (ECMC)\cite{bernard_2009}.  The equilibrium state obtained by ECMC is exactly equivalent to that obtained via Event-Driven Molecular Dynamics (EDMD)~\cite{isobe_1999} but is reached much faster than with conventional MC or EDMD, especially in large and dense systems. (For a comparison of CPU times see  Refs.~\cite{engel_2013,isobe_2015}).

The system is prepared for each packing fraction, $\nu=N\pi (x_0\sigma_0^2+x_1\sigma_1^2)/(4A)$, initially in equilibrium by performing long runs with ECMC for up to ${\cal O}(10^{13})$ collisions.  After equilibration, production runs for equilibrium dynamics are done using EDMD. Particle numbers are either $N=32 \times 32$ or $N=64 \times 64$, and packing fractions vary from $\nu=0.30$ to $\nu = 0.80$. The units are set by the mass $m$, the effective diameter $\sigma_*$ \cite{eff_d}, and energy $1/\beta$.

\begin{figure}[h!]
\begin{center}
\includegraphics[width=\columnwidth]{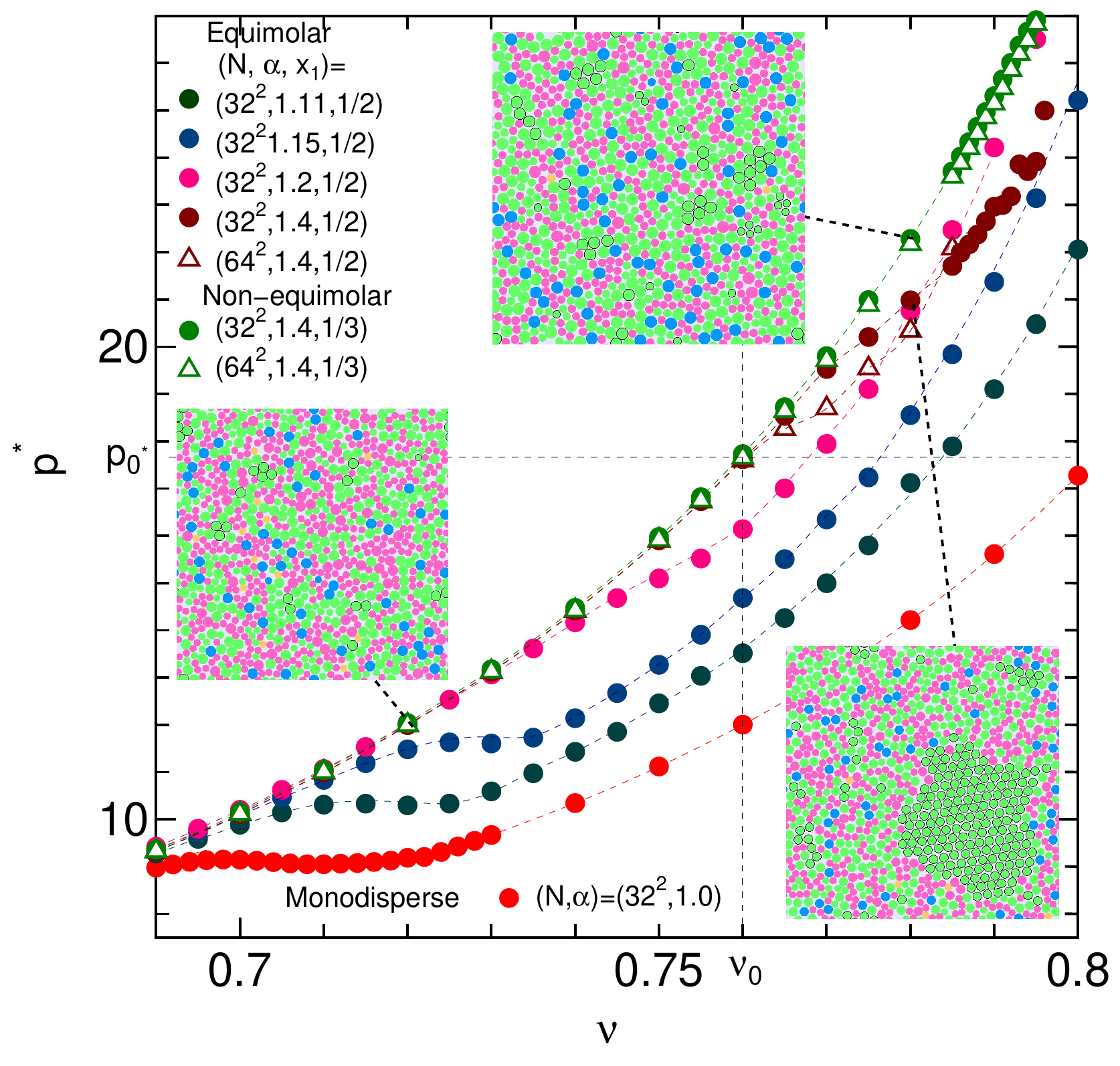}
\caption{
Phase diagram of hard disk binary mixtures~\cite{isobe_2016} in terms of the EOS.  Insets show examples of equilibrium configurations: for the non-equilmolar mixture with $(N, \alpha, x_{1})=(32 \times 32, 1.4, 1/3)$ we show a typical liquid configuration at $\nu = 0.720$, and a typical supercompressed configuration at $\nu = 0.780$.  In contrast, for the equimolar mixture $(N, \alpha, x_{1})=(32 \times 32, 1.4, 1/2)$ at $\nu = 0.780$ typical configurations show micro-crystallization of $40\%$ large disks immersed in the amorphous phase.
Note that when $x_1$ is decreased below $x_1=1/3$, micro-crystallization of small disks emerges.
Disks are colored by the number of nearest neighbours detected by a 2D version of the SANN algorithm~\cite{meel_2012} as 4(orange), 5(pink), 6(green), 7(blue) and 8(dark blue).  Disks belonging to a crystal cluster~\cite{cluster} are indicated also by a black perimeter.
}
\label{fig:2D_BI_EOS}
\end{center}
\end{figure}

Figure~\ref{fig:2D_BI_EOS} shows the phase diagram, via the equation of state (EOS), of the binary mixture hard disk system as generated with ECMC and EDMD, for size ratio $\alpha=1.0, 1.11, 1.15, 1.2, 1.4$ for an equimolar mixture, $x_1=1/2$, and a non-equimolar mixture, $x_1=1/3$ (see also Ref.~\cite{isobe_2016} for simulation details).  The reduced pressure $p^*$ in terms of the packing fraction $\nu$ is calculated by the virial form of collisions in EDMD~\cite{erpenbeck_1977}. In the liquid regions ($\nu < 0.69$), the pressure for different values of the parameters coincides with the universal liquid branch curve. Above $\nu > 0.70$, a phase transition occurs in the monodisperse system ($\alpha=1$)~\cite{alder_1962, bernard_2011}.  With increasing $\alpha$, the coexistence region ($0.70<\nu<0.72$) between liquid and crystal 
gradually shrinks and shifts toward higher densities and pressures, almost disappearing at $\alpha > 1.3$.

Overall we find four phases: an amorphous liquid, a pure crystal (only in the monodisperse case $\alpha=1$), a mixed crystal~\cite{mixed}, and a crystal-amorphous composite phase.
For an equimolar binary mixture $(\alpha, x_1)=(1.4, 1/2)$, when $\nu \geq \nu_0$, with $\nu_0$ being an onset packing fraction, all final configurations show micro-crystallization of large disks immersed in the amorphous state instead of a pure amorphous state. This is a slow coarsening process in which a crystal cluster emerges spontaneously, and sufficiently long simulation times are required to observe it (order of $10^{12}$ to $10^{13}$ collisions) in ECMC~\cite{CAS}. (Our EOS is compatible with that found in Ref.~\cite{huerta_2012} using Metropolis MC, for times $\sim 4 \times 10^9$ trial moves in systems of size $N=400$, although our systems are larger and we simulate for much longer times.) Above $\nu \sim 0.8$ the relaxation time is too large to observe equilibrium behaviour in our simulations. 

It is worth remarking that the equimolar mixture at $\alpha \sim 1.3$ was believed not to crystallize or demix \cite{speedy_1999,frobose_2000}, thus being studied extensively as a two-dimensional glass former. Subsequent work showed that in the supercompressed region this system had no ideal glass transition~\cite{santen_2000,donev_2006} and that true equilibrium was a stable crystal-amorphous composite (as in binary equimolar soft disks~\cite{perera_1999,cooper_2006}).  Previous simulations had not been able to observe spontaneous crystallization since the coexistence is difficult to realize directly due to surface tension between phases in small systems, and since diffusion of large disks towards nucleation is kinetically strongly suppressed~\cite{donev_2006,donev_2007}.

Typical configurations are shown as insets in Fig.~\ref{fig:2D_BI_EOS}.
For a non-equimolar binary mixture at $(\alpha, x_1)=(1.4, 1/3)$, a smooth continuous crossover from the normal liquid ($p^* < p^*_0$) to the supercompressed one ($p^* > p^*_0$) is observed, where $p^*_0 \approx 17.66$.
Above the freezing point of large disks ($p_f \approx 19.7$), the pressure in the equimolar case is lower than that of the non-equimolar due to microcrystallization.  Partial freezing of large hard disks above the freezing packing fraction, $\nu_f \approx 0.77$, also gives rise to a size dependence of pressure due to finite size effects on the surface tension of micro-crystallization.

The equilibrium phase diagram of Fig.~\ref{fig:2D_BI_EOS} suggests that $(\alpha, x_1)=(1.4, 1/3)$ is a suitable system to study features associated with glassy slowing down, as amorphous supercooled states can be prepared in equilibrium at supercompressed conditions of $p^* > p^*_0$.  Hereafter, we focus on the model with $(N, \alpha, x_1)=(64 \times 64, 1.4, 1/3)$ to investigate the accuracy of DF predictions to the dynamics of the hard disk binary mixture. 

\begin{figure}[h!]
\begin{center}
\includegraphics[width=\columnwidth]{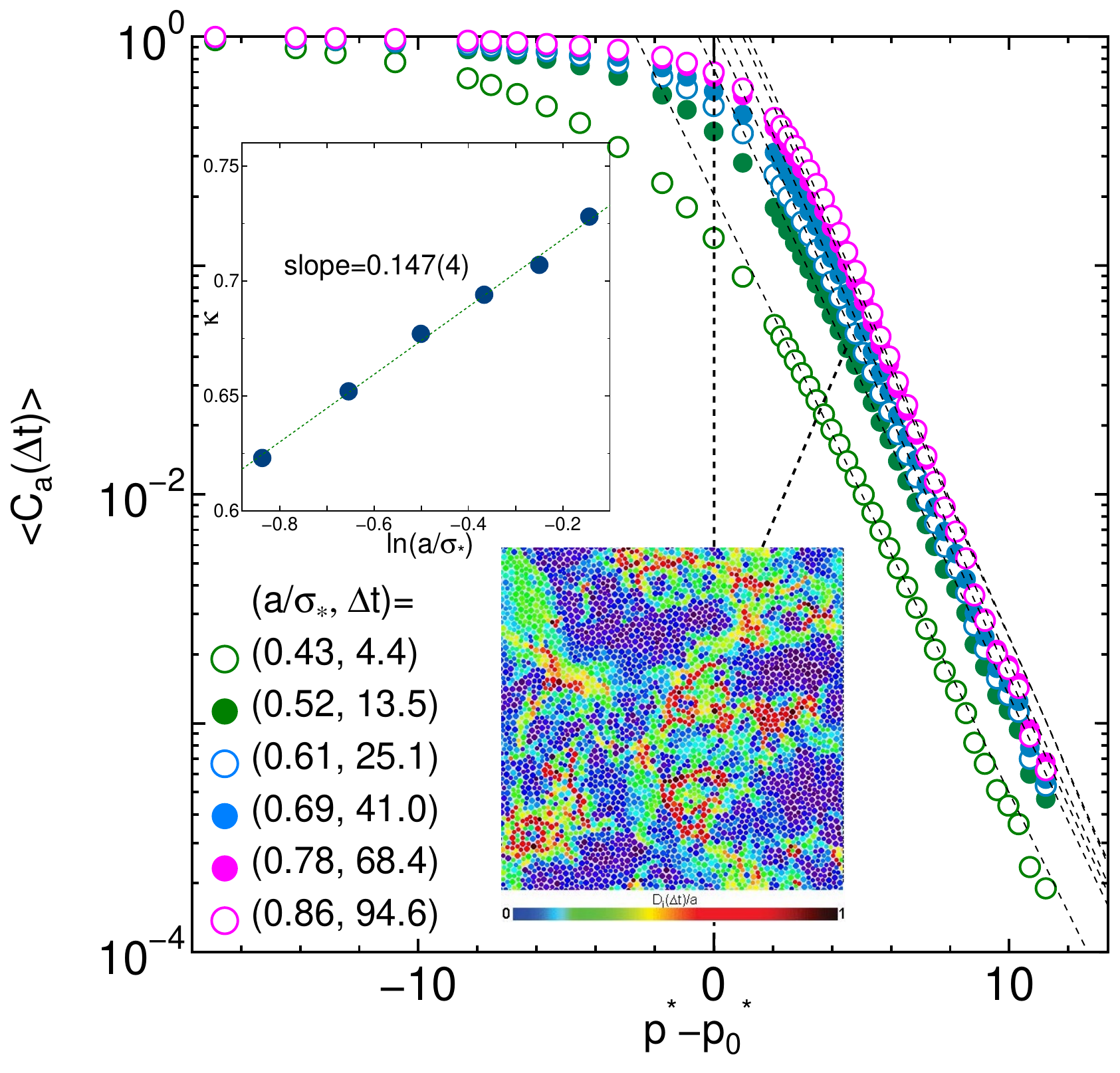}
\caption{
Averaged excitation indicator $\left<C_a (\Delta t)\right>$ as a function of pressure $p^*-p_0^*$, for several values of $(a, \Delta t)$ used to define an excitation, for the system with $(N,\alpha, x_1)=(64 \times 64,1.4,1/3)$. The inset shows the dependence with $a$ of the  parameter $\kappa_{a}$ obtained from fitting the data with  \er{eqn:edf1}. We also show a typical realisation of the displacement field $D_i(\Delta t)$ for the choice $(a/\sigma_*, \Delta t)=(0.52, 13.5)$ that illustrates dynamic heterogeneity in the system at the shown supercompressed conditions. 
}
\label{fig:H-p}
\end{center}
\end{figure}

We consider first the identification of excitations.  We follow the procedure of Ref.\ \cite{Keys2011} (see also \cite{Speck2012}).  We can quantify the overall density of excitations with the indicator function
\begin{equation}
C_a( \Delta t)=\frac{1}{N} \sum_{i=1}^N \theta(|\overline{{\bf r}_i}(\Delta t )-\overline{{\bf r}_i}(0)|-a) ,
\label{eqn:excitation}
\end{equation}
where $\theta$ is the Heaviside step function. In \er{eqn:excitation} excitations are associated with displacements of lengthscale $a$ (of the order of a particle diameter) that persist for a time $\Delta t$ \cite{Keys2011}.  To get rid of uninteresting short scale motion we consider positions averaged over a short timescale, $\overline{{\bf r}_i}(t)=\delta t^{-1} \int_{0}^{\delta t} {\bf r}_i(t+t') dt'$, with $\delta t$ large enough to suppress short scale fluctuations but still much shorter than relevant hopping times leading to structural rearrangements \cite{deltat}, cf.\ Ref.~\cite{Keys2011}.   
On average, the indicator \er{eqn:excitation} will grow linearly with $\Delta t$ as long as this {\em commitment time} is larger than the minimal time associated with a persistent displacement (or ``instanton time'' \cite{Keys2011}).  This means that we can extract the average concentration of excitations $c_{a}$ from the average of \er{eqn:excitation}, 
\begin{equation}
\langle C_a( \Delta t) \rangle \approx \Delta t \, c_{a} .
\label{eqn:ca}
\end{equation}
To extract the scaling of $c_{a}$ with pressure, cf.\ \er{eqn:edf1}, we consider fixed values of the commitment time for varying pressure.  Figure~\ref{fig:H-p} shows $\langle C_a( \Delta t) \rangle$ as a function of $p^*-p^{*}_{0}$ in the system with parameters $(N,\nu,\alpha,x_1)=(64 \times 64,0.780,1.4,1/3)$~\cite{Deltat}.
In the supercompressed regime , $p^* > p^*_0$, $\left<C_a\right>$ decays exponentially over almost three orders of magnitude 
with pressure, in agreement with the prediction of \er{eqn:edf1}.  This behaviour is systematic for a range of values of $a$ and $\Delta t$, suggesting that excitations can be identified robustly. 
Furthermore, the inset to Fig.~\ref{fig:H-p} shows that the rate of decay with pressure in the exponential function, $\kappa_{a}$ scales logarithmically with the lengthscale $a$, as predicted by \er{eqn:edf2}, with each $\kappa_{a}$ independently obtained from the exponential fits for $p^* > p^*_0$ of the main panel.

In Fig.~\ref{fig:H-p} we also give an illustration of the spatial distribution of excitations, whose facilitated motion gives rise to dynamic heterogeneity.  In the figure we plot an instance of the spatial displacement field $D_i(\Delta t) = |\overline{{\bf r}_i}(\Delta t)-\overline{{\bf r}_i}(0)|$ for all particles.  Mobile particles, $D_i(\Delta t) > a$, are coloured dark red, while immobile ones, $D_i(\Delta t) = 0$, dark blue, with disks with intermediate values, $0 < D_i(\Delta t)/a < 1$, coloured with a scale between these.

\begin{figure}[ht]
\begin{center}
\includegraphics[width=\columnwidth]{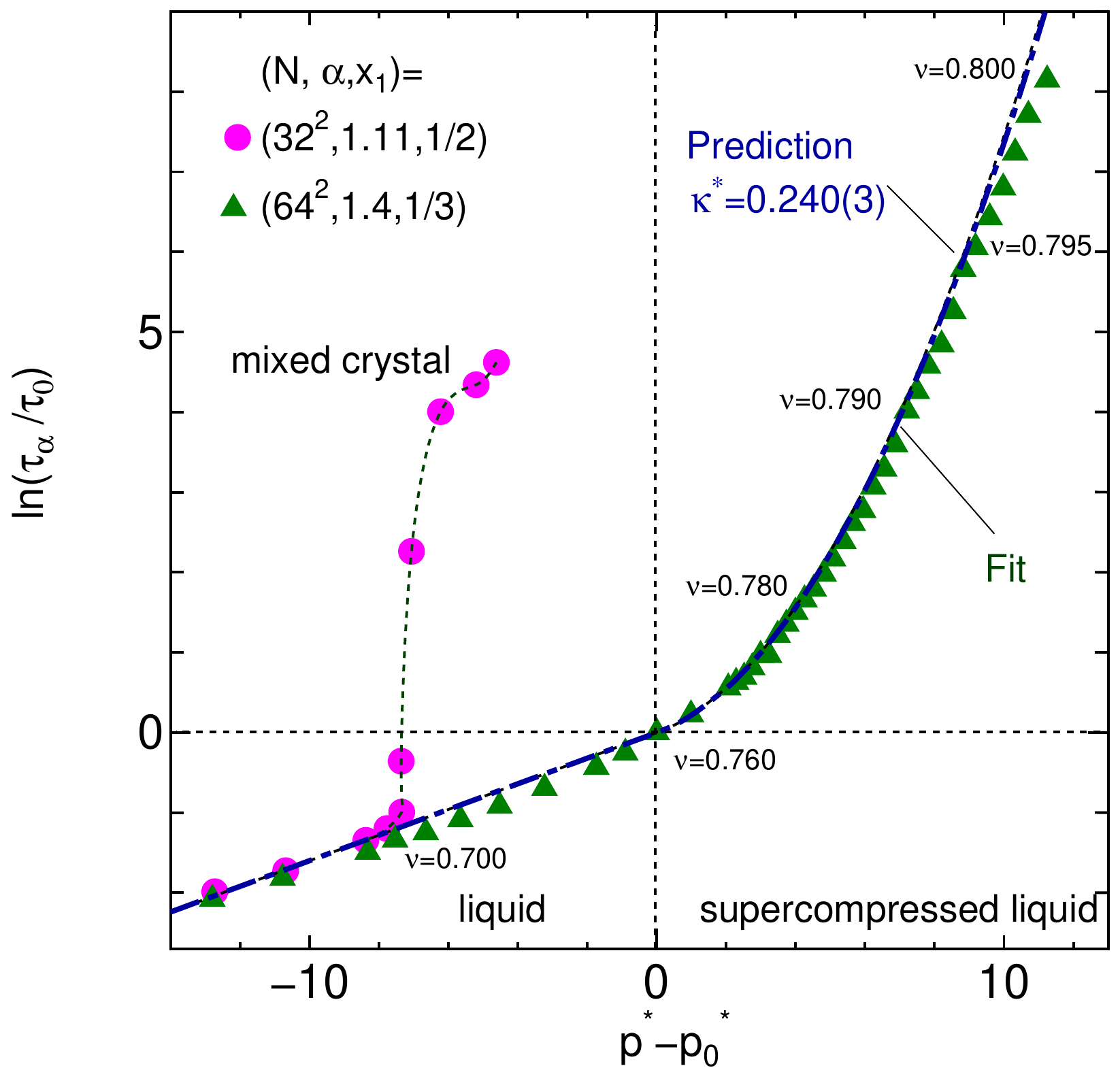}
\caption{
Structural relaxation time $\ln{(\tau_\alpha/\tau_0)}$ as a function of $p^*-p_0^*$ for the non-equimolar mixture $(N, \alpha, x_1)=(64 \times 64, 1.4, 1/3)$ as extracted numerically (symbols).  We also show a fit with \er{eqn:tau} (dashed line; fitting parameters are: $p_0^*=17.6568(4)$, $\tau_0=3.18$, $\kappa=0.241(4)$, $\lambda=0.169(4)$). 
We also show the timescale that would be predicted by taking $\kappa$ from the excitation data and using \er{eqn:tau} (dot-dashed). 
For comparison with the supercompressed case we show the data for the equimolar mixture $(N, \alpha, x_1)=(32 \times 32, 1.11, 1/2)$ which crystallises at these conditions. 
}
\label{fig:taua_p}
\end{center}
\end{figure}

We now consider the behaviour of the overall relaxation time in the supercompressed regime.  We estimate $\tau_{\alpha}$ from the self-intermediate scattering function  $F_s(k,t)= N^{-1} \sum_{i=1}^N \left< \exp{[-i{\bf k}\cdot ({\bf r}_i(t)-{\bf r}_i(0))]}\right>$ at wave vector  
$k^*=2\pi/\sigma_*$.  Specifically, we extract $\tau_{\alpha}$ from  $F_s(k^{*},\tau_{\alpha}) = 10^{-1}$.  Fig. \ref{fig:taua_p} shows the obtained $\tau_\alpha$, in units of $\tau_{0}=\tau_\alpha(p^{*}_{0})$,  versus $p^*$.  For high pressures we expect the relaxation time to obey \er{eqn:edf3}.  In order to fit both high and low pressures we use the form
\begin{equation}
\ln \left(\frac{\tau_\alpha}{\tau_0}\right) = 
\left\{
\begin{array}{lcl}
e^{\kappa^2 (p^*-p^*_0)^2+\lambda(p^*-p^*_0)}
&& p^* \geq p^*_0 \\ 
e^{\lambda(p^*-p^*_0)}
&& p^* < p^*_0 \\ 
\end{array}
\right.
\label{eqn:tau}
\end{equation}
where $\kappa$ and $\lambda$ are fitting parameters.  \er{eqn:tau} interpolates from the low pressure liquid regime to the high pressure supercompressed regime.  \er{eqn:tau} (dashed line) fits well the data (symbols), as shown for the system with the non-equimolar mixture $(N, \alpha, x_1)=(64 \times 64, 1.4, 1/3)$, over a range of timescales spanning four orders of magnitude.  Note that while for $p^* < p^*_0$ the relaxation time behaves in an ``Arrhenius'' fashion (in the sense that it appears exponential in $p^{*}$), for $p^* > p^*_0$ the behaviour is ``super-Arrhenius''.  The crossover to super-Arrhenius behaviour indicates dominance of heterogeneous dynamics.  From Figs.~\ref{fig:2D_BI_EOS}-\ref{fig:taua_p}, we identify the onset pressure $p^*_0$ as the point at which this crossover takes place. 
That the overall relaxation rate can be accounted for with a parabolic law is another indication of the validity of the DF approach to interpret the phenomenology of glass forming systems with hard core interactions~\cite{relaxtime_nu}.

The change from Arrhenius to super-Arrhenius behaviour 
in the non-equimolar mixture occurs in the absence of a structural change to either a mixed crystal or due to micro-crystallization.  
To highlight this, we also show in Fig.~\ref{fig:taua_p} the structural relaxation of the equimolar mixture $(N, \alpha, x_1)=(32 \times 32, 1.11, 1/2)$ which does crystallise. In this case, the time suddenly increases around $p^* \sim 10$ due to the transition from liquid to mixed crystal, see Fig.\ \ref{fig:2D_BI_EOS}, in a manner which is quite distinct to that of the supercompressed system.  Furthermore, in contrast to the equimolar case at $\alpha=1.4$, whose freezing packing fraction is $\nu_f \approx 0.77$, for the non-equimolar mixture 
there is no freezing point in terms of micro-crystallization at least up to $\nu=0.80$, see Fig.~\ref{fig:2D_BI_EOS}.

The results for the density of excitations, Fig.~\ref{fig:H-p}, and for the relaxation time, Fig.\ \ref{fig:taua_p}, provide two independent estimates of the energy scales from \ers{eqn:edf1}{eqn:edf2} and \er{eqn:edf3}, respectively.  From DF we would expect that the ratio $\kappa/\kappa^*$, where $\kappa^*=\kappa_{\sigma_*} \sqrt{\gamma/d_f}$, should be $\kappa/\kappa^* \approx 1$, where $d_f = 1.8 \sim 1.9$ is the fractal dimension associated to hierarchical facilitation estimated numerically in Ref.\ \cite{Keys2011} for systems in two dimensions.  In the non-equimolar mixture studied, $\kappa \sim 0.241(4)$ from the relaxation time, cf. Fig.~\ref{fig:taua_p}, while $\kappa^* \sim 0.240(3)$ (in case of $d_f=1.9$) from the excitation data, since $(\kappa_{\sigma_*}, \gamma)=(0.748(2), 0.196(5))$, cf.\ Fig.~\ref{fig:H-p}).  The ratio is then 
\color{black}
 $\kappa/\kappa^* = 1.00(2)$ 
\color{black}
 which is close to unity.  This means that the relaxation time for pressures above the onset could have been predicted directly from the excitation concentration data using $\kappa^*$ in the parabolic law \er{eqn:edf3}.
The accuracy of this predicted fit is shown in Fig.~\ref{fig:taua_p} (dashed-dotted line)~\cite{high_pf}.

In summary, we have performed extensive numerical simulations of binary hard disk mixtures to study their dynamical properties as a function of pressure.  We have shown that the basic predictions of DF theory seem to hold: (i) localised effective excitations, distributed randomly in equilibrium, giving rise to facilitated relaxation; and (ii) average relaxation times grow with increasing compression as the exponential of a quadratic function of pressure.  A key aspect of our study is the access to true equilibrated systems at all conditions studied by means of novel event-driven algorithms.  Being able to precisely identify the four existing equilibrium phases allowed us to study relaxation in the true supercompressed regime by considering non-equimolar mixtures that do not phase separate or undergo micro crystallisation at the high pressures considered. 

As for systems with soft interactions \cite{Keys2011, Speck2012}, we have shown here that the basic aspects of DF theory seem to apply to systems with hard-core potentials in the regime where relaxation is slow, cooperative and heterogeneous.  In future work, it would be important to test prediction (iii) relating to the existence of trajectory space transitions to inactive dynamical phases.  Interesting questions relate to whether such inactive phases 
display structural features associated to disordered packings of hard objects, such as hyperuniformity~\cite{Donev2005, Zachary2011}, and to the connection between non-ergodic inactive states and glasses obtained by standard compression, cf.\ \cite{Keys2015}. 

\begin{acknowledgments}
MI was supported by JSPS KAKENHI Grant Number JP26400389.
Part of the computations were performed using the facilities of the Supercomputer Center, ISSP, Univ. of Tokyo.
DC was supported in part by a grant from the U.S. National Science Foundation.  ASK was supported by the Director, Office of Science, Office of Basic Energy Sci- ences, and by the Division of Chemical Sciences, Geosciences, and Biosciences of the U.S. Department of Energy at LBNL, by the Laboratory Directed Research and Development Program at LBNL under Contract No. DE-AC02-05CH11231.
JPG was supported by EPSRC Grant No.\ EP/K01773X/1.
\end{acknowledgments}


\begin{thebibliography}{99}
\bibitem{eff_d} The effective diameter $\sigma_*$ can be directly estimated by averaging diameter $\left< \sigma_{ij} \right>$ for collision pair $(i,j)$ during the EDMD simulation. In the equimolar ($x_1=1/2$) and non-equimolar ($x_1=1/3$) cases, $\sigma_*/\sigma_0$ are estimated as $1.2248(5)$ and $1.1556(6)$, respectively.
\bibitem{mixed} The mixed crystal is one where particles are located almost on the triangular lattice, as in the monodisperse case, but the six neighbors of a disk include both large and small particles.
\bibitem{CAS} Independent runs of Event-Driven MD instead of Event-Chain MC confirm that the final equilibrium states are same. The longest simulation length of Event-Driven MD is ${\cal O}(10^{12})$, which are $10^2 \sim 10^3$ longer than that of past literature.
\bibitem{cluster} To detect a disk belonging to crystal cluster we proceed as follows: if a large tagged disk has six nearest neighbours, then we check the character of these neighbours; if there are more than four large disks which also have six large nearest neighbours, it is categorized as part of a crystal cluster.
\bibitem{deltat}
Typical trajectories of a tagged disk at supercompressed conditions show that a coarse-graining time of $\delta t=4.0$, which is about $100$ times larger than the mean free time around the transition point in the monodisperse hard disk system, is enough to suppress vibrational fluctuation and is also sufficiently small compared to timescales of disk diffusion from the their local trapping cage.
\bibitem{Deltat}
The characteristic time scales $\Delta t$ of the excitations for each $a$ were estimated by preliminary independent runs.  In the production runs the averaged indicator function $\left<C_a (\Delta t)\right>$ was calculated for each pressure. This allowed to obtain $\kappa_a$ from a fit to an exponential in the pressure, above the crossover pressure $p_0^*$.
\bibitem{relaxtime_nu}
In hard sphere systems, pressure is the appropriate intensive control variable.  Nevertheless, the relaxation time is often
quoted as a function of the packing fraction. In our case this can be done by relating $p^{*}$ to $\nu$ from the phase diagram of Fig.\ 1. To the extent that one can neglect curvature in the $\nu$-$p^{*}$ plot, Eq.\ (\ref{eqn:tau}) will also yield a parabolic law for 
$\ln \tau_\alpha$ vs.\ $\nu$.
\bibitem{high_pf}
At high packing fractions, $\nu \sim 0.800$, there is a systematic deviation from the DF prediction.  Note that there is also a slight deviation from the exponential fit in Fig.~2 in the range $\nu \geq 0.795$ ($p^*-p_0^* \geq 9.2$).  This regime is where equilibration is most difficult in simulations, even with the efficient algorithm employed here; these deviations may be an indication of incipient lack of equilibration (which may be more evident in dynamical observables than in the static ones that define the phase diagram of Fig.1).
\end{thebibliography}
\end{document}